\documentclass[aps,prc,twocolumn,showpacs,superscriptaddress,groupedaddress]{revtex4} 

\usepackage{graphicx} 
\usepackage{amsfonts}
\usepackage{booktabs}
\usepackage{siunitx}
\usepackage{xcolor}
\usepackage{dcolumn}
\usepackage{multirow}
\usepackage{hyphenat}
\usepackage{ulem}

\begin{document}

\title{Position uncertainties of AGATA pulse-shape analysis estimated via the bootstrapping method}
\author{M.~Siciliano}
	\thanks{Present address: Argonne National Laboratory, Lemont (IL), United States}
	\affiliation{Irfu/DPhN, CEA, Universit\'e Paris-Saclay, Gif-sur-Yvette, France}
\author{J.~Ljungvall}
	\affiliation{IJC Lab, CNRS/IN2P3, Universit\'e Paris-Saclay, Orsay, France}
\author{A.~Goasduff}
	\affiliation{Dipartimento di Fisica e Astronomia, Universit\'{a} di Padova, Padova, Italy}
	\affiliation{INFN, Sezione di Padova, Padova, Italy}
\author{A.~Lopez-Martens}
	\affiliation{IJC Lab, CNRS/IN2P3, Universit\'e Paris-Saclay, Orsay, France}
\author{M.~Zieli\'{n}ska}
	\affiliation{Irfu/DPhN, CEA, Universit\'e Paris-Saclay, Gif-sur-Yvette, France}
\author{AGATA and OASIS collaborations}

\begin{abstract}
\noindent
The unprecedented capabilities of state-of-the-art segmented germanium-detector arrays, such as AGATA and GRETA, derive from the possibility of performing pulse-shape analysis. 
The comparison of the net- and transient-charge signals with databases via grid-search methods allows the identification of the $\gamma$-ray interaction points within the segment volume. 
Their precise determination is crucial for the subsequent reconstruction of the $\gamma$-ray paths within the array via tracking algorithms, and hence the performance of the spectrometer. 
In this paper the position uncertainty of the deduced interaction point is investigated using the bootstrapping technique applied to $^{60}$Co radioactive-source data.
General features of the extracted position uncertainty are discussed as well as its dependence on various quantities, e.g. the deposited energy, the number of firing segments and the segment geometry.
\end{abstract}

\maketitle

\section{Introduction}
\label{sec:intro}

For more than 50 years, High-Purity Germanium (HPGe) detectors have been at the heart of powerful $\gamma$-ray spectrometers aiming at unraveling the complex structure of the atomic nucleus. 
Recent developments in detector technology, data acquisition and processing led to a refinement of the detection techniques, making it possible to deduce position information from the signal shapes of semiconductor detectors. 
Electrical segmentation of a HPGe detector not only allows us to improve the angular granularity and the position sensitivity at the level of the physical segmentation, but it permits to deduce precise information on the $\gamma$-ray interaction points from a comparison of signals induced in the neighboring segments.
In order to perform such a pulse-shape analysis, segment signals need to be acquired with high resolution, high bandwidth and high sampling frequency over a meaningful time period. 
Thanks to these technological developments, new generation HPGe arrays, e.g. AGATA~\cite{akkoyun2012agata} and GRETA~\cite{paschalis2013GRETA}, are being constructed, with a capability to determine the position of charge generation inside the detector volume from the pulse-shape information.

The \textit{Advanced GAmma Tracking Array}~\cite{akkoyun2012agata,AGATA2020whitebook} collaboration aims at the construction of a European $4\pi$ $\gamma$-ray tracking array, which will constitute an unparalleled tool for investigation of nuclear structure of both stable and exotic species via high-resolution $\gamma$-ray spectroscopy. 
After the first campaign of the AGATA demonstrator in Legnaro National Laboratories (LNL)~\cite{AGATA@LNL2011}, the array moved to GSI~\cite{AGATA@GSI2014,Lalovic2016258} to take advantage of the radioactive-ion beams during the PreSPEC campaign, and then to GANIL~\cite{AGATA@GANIL2017} where it has reached a nearly $1\pi$ configuration with 42 HPGe crystals. 
In the near future, AGATA will return to LNL for an experimental campaign with stable beams delivered by the XTU-TANDEM/ALPI/PIAVE accelerator complex and with exotic beams produced by the SPES facility. 
The excellent efficiency and energy resolution of the array, together with its outstanding position resolution, in particular when coupled with $\gamma$-ray tracking algorithms, contributed to the rich physics output of these experimental campaigns. 

In order to identify the position where a $\gamma$-ray has interacted inside an AGATA segmented detector, the pulse-shape analysis (PSA) is performed by comparing the experimental signals with a basis of position-dependent pul\hyp{}ses. 
The reference pulses are presently generated via simulations using the AGATA Detector Library (ADL)~\cite{bruyneel2016ADL}.
This procedure requires very accurate information on the properties of each individual detector, such as the geometry, the space-charge distribution, crystal-axis orientation, cross-talk properties and the response function of the electronics~\cite{BRUYNEEL2006764,BRUYNEEL2009196,BIRKENBACH2011176,BRUYNEEL201192,wiens2013} and yields a set of simulated signals associated with every point of a 2-mm step grid. 
Moreover, several groups have been working on the characterization of AGATA detectors with scanning tables~\cite{CRESPI2008440,Dimmock2009characterisation,HA2013123,Goel2013,Ginsz2015,HERNANDEZPRIETO201698,HABERMANN201724,deCanditiis2020}, in order to produce a realistic basis of pulses, independent from simulation assumptions and free of a potential bias introduced by boundary conditions. 
Following the procedures outlined in Refs.~\cite{PhysRevC.62.024614,DESESQUELLES2011324,DESESQUELLES2013198}, a new experimental basis was recently developed using radioactive-source data~\cite{Li2018}. 
Such pulses can be extracted \textit{ad-hoc} for each experiment during the calibration and they take into account the real experimental conditions, including alterations of the signals due to the electronic response or specific features of individual detectors. 
However, a slightly worse performance observed for such an experimentally deduced basis~\cite{Li2018} suggests that they are less accurate than the ADL bases.

During the PSA process, the comparison of measured ($m$) and simulated ($s$) signals is done by minimizing the following figure of merit (FoM):

\begin{equation}
    FoM = \sum_{j, t_i} |A_{j,m}(t_i) -A_{j,s}(t_i)|^p \, ,
\label{eq:FoM}
\end{equation}
where $A_{j,m}(t_i)$ and $A_{j,s}(t_i)$ are the amplitudes at the $t_i$-th sample of the signal from segment $j$.
In the sum only the 56-samples traces for the segment measuring the net charge, for the neighboring ones (typically 4-5 segments) and the core are considered. 
As shown in Figure~\ref{fig:Trace}, such short traces include all features of the signal that are necessary for the analysis of the pulse shape: the baseline (8 samples), the rising part of the signal ($\approx 20$ samples) and its saturation. 
In the FoM of Equation~\ref{eq:FoM}, the exponent $p$ depends on the adopted metric. 
Studies of the optimization of the Doppler correction led to the adopted value $p=0.3$~\cite{RecchiPHD}. 

Recently, a different definition of the figure of merit has been proposed~\cite{lewandowski2019PSA}: 

\begin{equation}
    FoM = \sum_{j} w_j \sum_{t_i} |A_{j,m}(t_i) -A_{j,s}(t_i)|^{p_j} \,
    \label{eq:lewandowski1}
\end{equation}
where the weighting coefficient $w_j$ and the metric exponent $p_j$ have different values for the transient signals and for those of the hit segment and the core. 
This alternative definition enables taking into account the different position sensitivity of the two types of signals. 
In fact, the signals of the hit segment and of the core change rapidly with the radial position of the interaction point, while the transient signals provide a more precise information on the $\phi$ and $Z$ coordinates. 
The proposed FoM gives better performance of the PSA procedure, but it is worth noting that the best results were obtained assuming crystal and electronics parameters (e.g. charge carrier mobilities, electronics transfer function) different from those measured and commonly used. 

Since PSA does not make use of the $\chi^2$ minimization to identify the $\gamma$-ray interaction points, standard statistical procedures cannot be used to infer the position uncertainty. 
This information is crucial for future developments of the tracking algorithms, as it would allow assigning proper weights to the determined interaction points when reconstructing the $\gamma$-ray path inside a HPGe array. 
In addition, the energy and segment dependence of the position uncertainty may be used to optimize the PSA algorithms, e.g. by preventing them from looking for the interaction point with a precision better than what can be reached, in order to reduce the computational load of the procedure. 
The PSA is currently the main factor limiting the counting rate that can be accepted by data acquisition and online analysis, and an improved throughput is therefore of great interest. 
The position resolution of AGATA detectors has been studied in multiple works via different techniques~\cite{RECCHIA200960,recchia2009position,soderstrom2011interaction,Steinbach2017}, with the resulting full width at half maximum (FWHM) of about 5 mm for an energy deposition of 1332.5 keV. 
A recent study based on the Doppler-correction capabilities of the AGATA-VAMOS++ experimental setup~\cite{LJUNGVALL2020163297} yielded an average value about 16\% larger than the previous estimations resulting from dedicated experiments. 
On the other hand, the position resolution estimated from experimental data does not correspond to the position resolution needed to optimize the performance of the $\gamma$-ray tracking. 
In the Orsay Forward Tracking algorithm~\cite{lopez2004gamma}, indeed, the position resolution enters as a parameter that was optimized on data simulated assuming a position resolution of 5 mm FWHM. 
However, this parameter has to be increased by a factor of 3~\cite{Li2018} when tracking experimental data, suggesting a worse position resolution. 
This effect may be related to a non-Gaussian distribution of the position uncertainties, which would lead to a larger value of the effective position resolution required by $\gamma$-ray tracking algorithms. 
However, the actual shape of the uncertainty distribution is not known.

In this context, this article presents an application of the bootstrapping technique to pulse-shape source data, providing an uncertainty estimate for $\gamma$-ray interaction positions in AGATA detectors for the commonly used PSA grid search. 
After discussing the differences between the \textit{Full} and \textit{Adaptive} grid search, the dependence  of the position uncertainty on various variables, such as the deposited energy, the number of firing segments within the same HPGe crystal and the segment geometry, is studied for the latter.

\section{Methods}
\label{sec:methods}

\begin{figure*}
    \centering
    \includegraphics[width=0.90\textwidth]{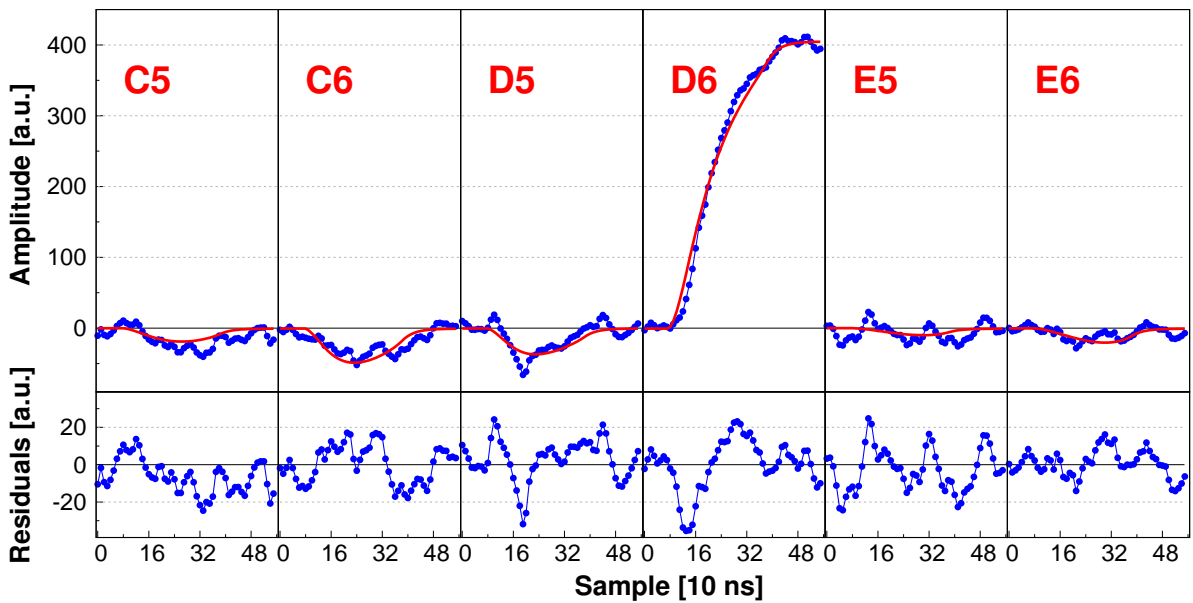}
    \caption{\label{fig:Trace}(Color online) (top) Comparison between a measured pulse (blue points and line) and the corresponding PSA-basis signal resulting from the grid search (red line). The traces are reported for the firing segment (D6) and those of the neighboring segments that are considered in the PSA procedure. (bottom) Residuals given by the sample-by-sample difference between the measured pulse and the basis signal.}
\end{figure*}

Bootstrapping is a statistical technique based on random sampling with replacement, used to estimate statistical properties of a distribution, such as its standard deviation, when the shape of the distribution is unknown~\cite{efron1986bootstrap,varian2005bootstrap}. 
This method can be used to construct hypothesis tests, in particular when parametric inference is impossible or impractical due to the complicated formulas it would require. 
The basic idea of bootstrapping is that inference about a distribution from sample data can be modeled by resampling the data and performing inference about the sample from the resampled data. 
Since the distribution is generated from the data itself and the investigated hypothesis (in this case: the entire PSA procedure), this statistical approach is completely independent from external assumptions or boundary conditions. 

In addition to its simplicity, the possibility to evaluate the position uncertainty along the three axes independently represents a great advantage of the bootstrapping technique with respect to the former studies. 
For instance, the Doppler-correction approach is unable to discriminate between different positions characterized by the same $\gamma$-ray emission angle, resulting in a possible underestimation of the position resolution.
Moreover, a great advantage of the bootstrapping technique is its wide applicability. 
In fact, contrary to other approaches that require specific experiments or complementary detectors, the discussed method can be applied to any kind of data. 
Since the measured traces are usually registered for AGATA in-beam experiments and radioactive-source runs, the position uncertainties can be evaluated for each data set independently by applying the bootstrapping technique during the PSA procedure.

The results presented in this manuscript are based on the data collected with a $^{60}$Co radioactive source, placed at the center of the AGATA array. 
Although simulations suggest that neutron damage has very minor influence on the PSA performance~\cite{DESCOVICH2005199}, two HPGe crystals have initially been considered: crystal C013 (in a good condition) and crystal B004 (with severe neutron damage). 
A standard indicator of neutron damage is the full width at tenth of maximum (FWTM). 
The FWTM/FWHM ratio for an ideal Gaussian is 1.8,  while the average FWTM/FWHM values measured at 1332.5 keV for C013 and B004 were around 1.9 and 4.2, respectively.
The data were collected during the AGATA experimental campaign at GANIL (for crystal C013 in June 2017, and for crystal B004 in February 2020) and the local-level processing preceding the PSA followed the standard procedures, described in detail in Ref.~\cite{LJUNGVALL2020163297}. 
The present study showed no noticeable difference between the PSA performance for C013 and B004, and thus only the results for C013 are presented in the following. 

\begin{figure}
    \centering
    \includegraphics[width=0.37\textwidth]{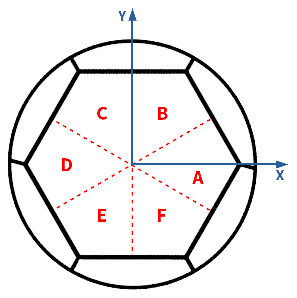}    
    \includegraphics[width=0.37\textwidth]{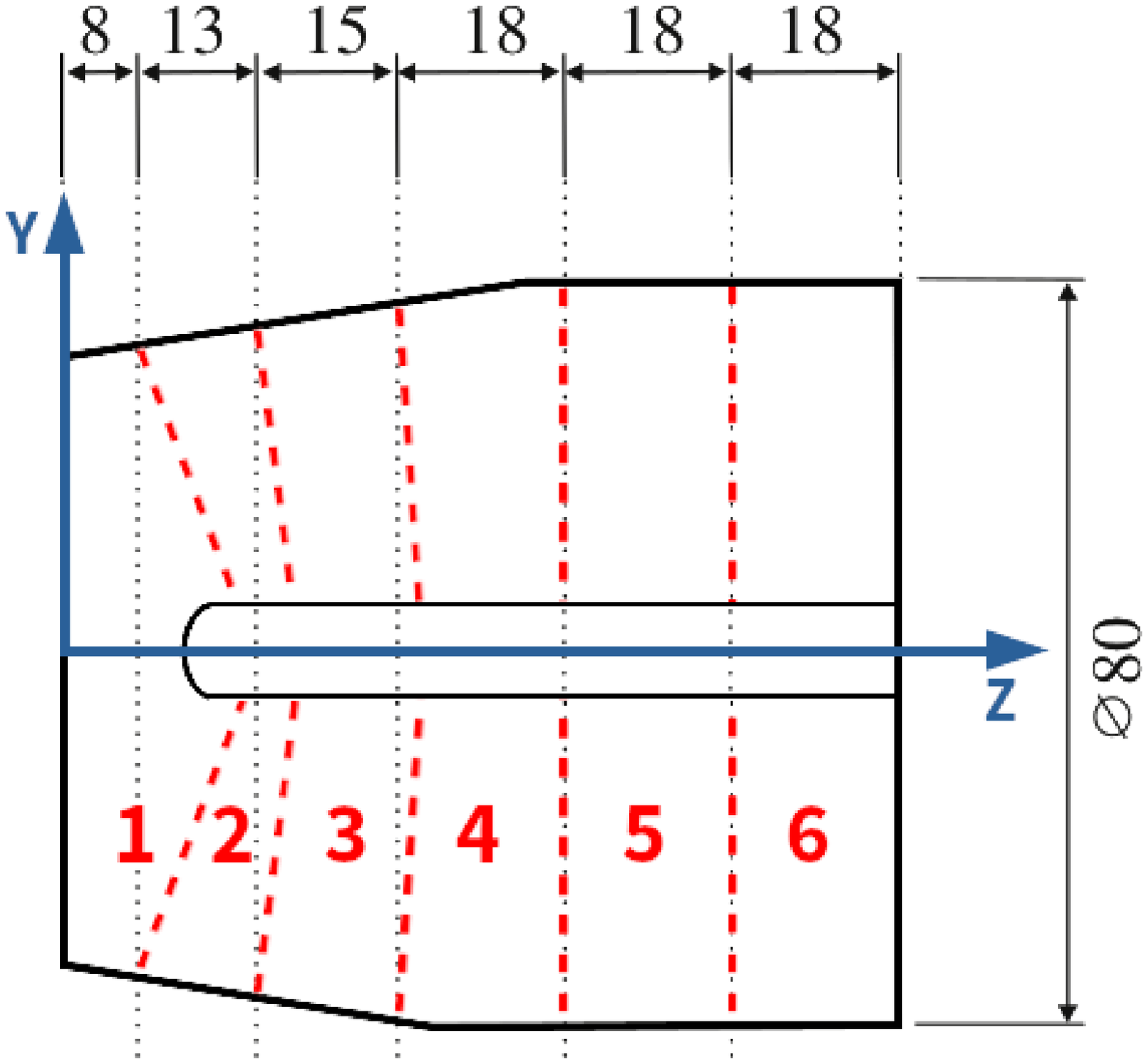}    
    \caption{\label{fig:FoR} (Color online) Sketch of an AGATA detector, showing the (top) front and (bottom) lateral view. Blue arrows represent the internal frame of reference of the PSA coordinates. Nominal dimensions of the segments (black dotted lines, arrows and labels) are expressed in millimeters and compared with the actual geometry (red dashed lines).  Figure adapted from Ref.~\cite{akkoyun2012agata}. }
\end{figure}

As shown in Figure~\ref{fig:Trace}, the grid search identifies the PSA basis signal that minimizes the FoM of Equation~\ref{eq:FoM}, providing the $\vec{r}_i \equiv (X,Y,Z)_i$ position vector ($X$, $Y$, and $Z$ refer to the intrinsic reference system of each AGATA crystal, as shown in Figure~\ref{fig:FoR}). 
From the comparison between the measured pulse and the energy-normalized basis, the residuals can be calculated. 
For each original event, a new population of pulses was generated by adding a random value of the residuals to the simulated pulse on a sample-by-sample basis. 
In order to avoid bias caused by a possible presence of systematic deviations between the measured traces and the basis, each segment is treated independently by considering only its own 56-samples set of residuals, which was randomly chosen without repetition. 
Moreover, in order to ensure the asymptotic convergence of the results, the resampled population has to be sufficiently large, which means that the bootstrapping procedure needs to be repeated multiple times. 
In the present study, 1000 new pulses were generated for every original pulse. 
This means that, despite its simplicity, the bootstrap procedure can be very demanding in terms of time and memory, in particular when the generated traces are written on disk. 
Therefore, an additional constraint was required in order to limit the number of iterations. 
To account for the energy dependence of position uncertainties discussed in Ref.~\cite{soderstrom2011interaction}, 8 different energy ranges have been defined between 8 keV and 2048 keV (see Figure~\ref{fig:Error-Energy}) and for each of them an upper threshold of $10^9$ bootstrapping events was imposed. 
This level of statistics was found sufficient to estimate the PSA uncertainty and its properties as a function of the deposited $\gamma$-ray energy. 
Finally, the ``bootstrap'' traces were processed by the PSA, providing a new set of $\vec{r}_{i,j} \equiv (X,Y,Z)_{i,j}$ position coordinates, where the index $j$ runs over all ``bootstrap'' traces. 
These coordinates were compared with those obtained from the original traces. 

\section{Results}
\label{sec:results}

The position fluctuations $\vec{\Delta_{i,j}} \equiv \vec{r}_i - \vec{r}_{i,j}$ are obtained for each coordinate ($X,Y,Z$) set and the uncertainties are defined as the standard deviation of their distributions. 
Information on the position uncertainty is deduced independently for the three coordinates by studying the distribution of the fluctuations along each axis, i.e. $\Delta_X$, $\Delta_Y$ and $\Delta_Z$ (for clarity, the index $i,j$ are dropped when the coordinate is explicitly given). 

\begin{figure*}
    \centering
    \includegraphics[width=0.42\textwidth]{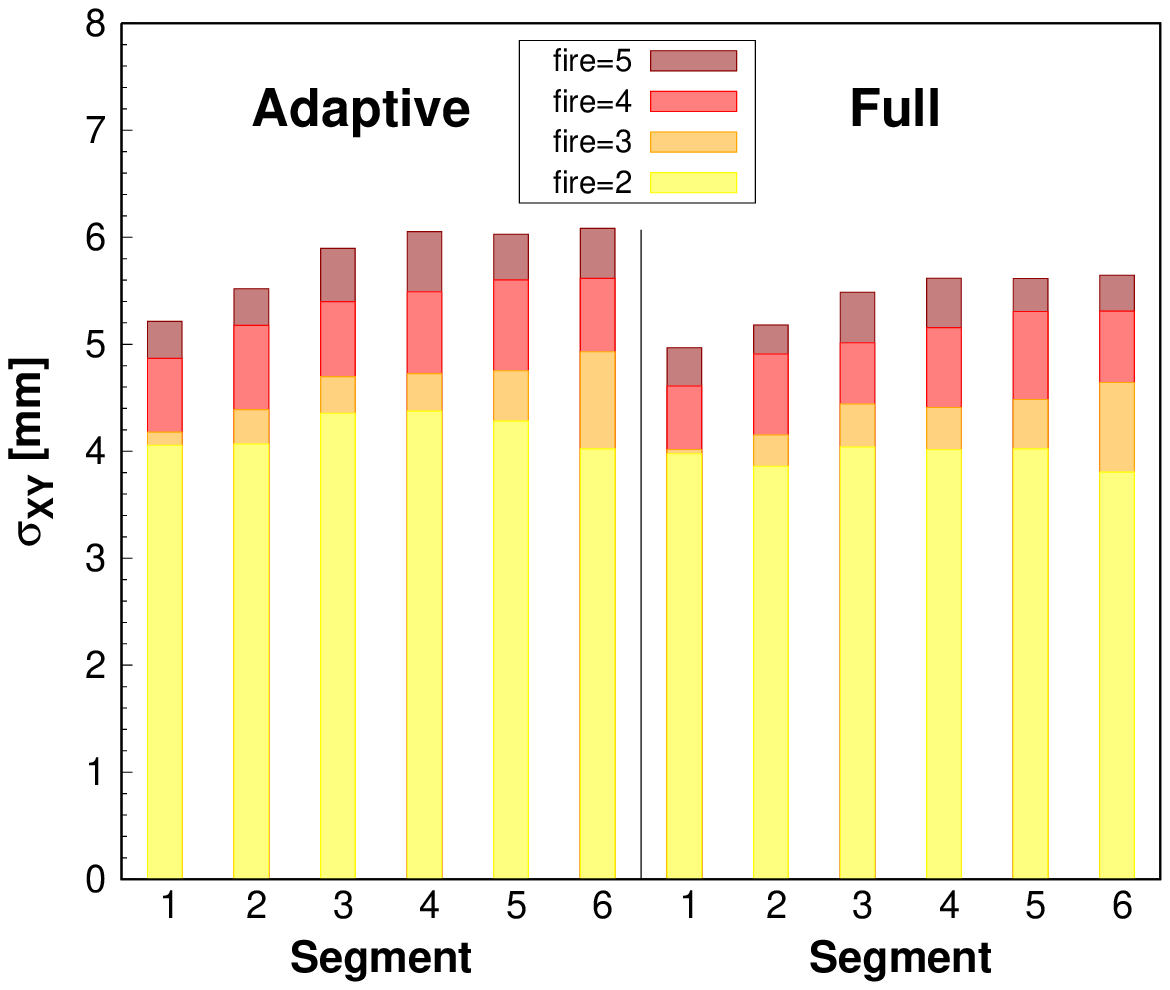}
    \hspace{5mm}
    \includegraphics[width=0.42\textwidth]{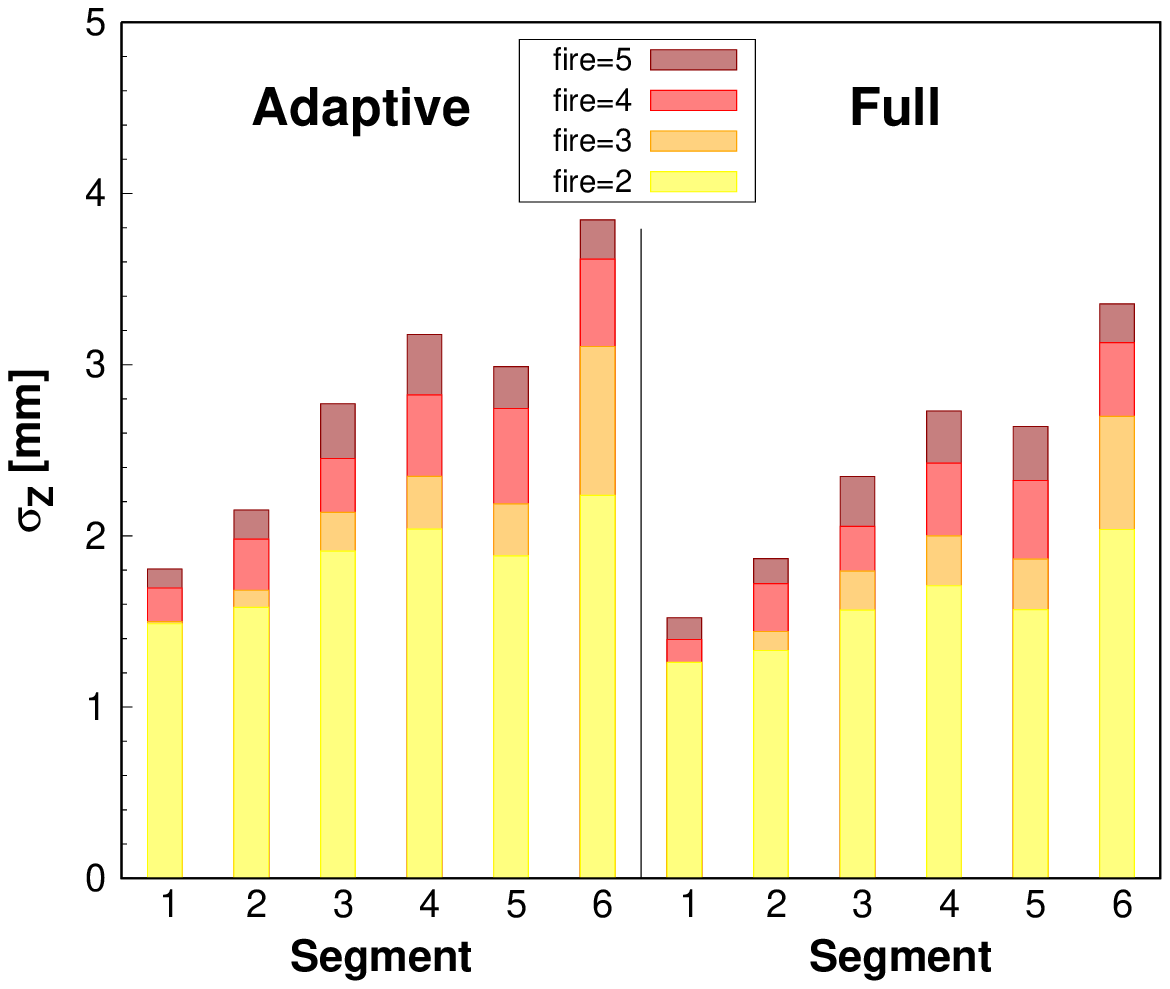}
    \caption{\label{fig:Error-Firing}(Color online) Position uncertainties as a function of the number of firing segments within the same crystal, for the (left) radial and (right) $Z$ directions and for the different segment geometries. The results are reported for the \textit{Adaptive} and \textit{Full} grid search of AGATA PSA.} 
\end{figure*}

The $\Delta_X$, $\Delta_Y$ and $\Delta_Z$ fluctuations were investigated as a function of the segment geometry and position. 
Taking into account the shape of the detector and the symmetry of the electric field (see Figure~\ref{fig:FoR}), one would expect the fluctuation distributions to be very similar for the $X$ and $Y$ directions. 
Such a symmetry was indeed observed and the variations of the  $\Delta_X$, $\Delta_Y$ and $\Delta_Z$ fluctuations were only due to the orientation of the segments with respect to the frame of reference. 
A radial fluctuation $\Delta_{XY} = \sqrt{\Delta_X^2 +\Delta_Y^2}$ was, therefore, introduced. 
As shown in the left panel of Figure~\ref{fig:Error-Firing}, the standard deviation $\sigma_{XY}$ remains rather constant and the small variations reflect the tapering of the HPGe crystals. 
In contrast, the uncertainty along the $Z$ axis ($\sigma_{Z}$) increases with the interaction depth (see right panel of Figure~\ref{fig:Error-Firing}). 
This trend, which is consistent with the results of Ref.~\cite[Fig.14]{soderstrom2011interaction}, may have two origins. 
On one hand, the segment height increases from the front to the back of the detector~\cite[Fig.4]{akkoyun2012agata}, so the range of the $\Delta_Z$ fluctuations increases as well, being less restrictive for the bottom segments. 
On the other hand, since the radioactive $\gamma$-ray source was placed at the center of the reaction chamber, the segments at the back of the crystal were reached predominantly by high-energy $\gamma$ rays, which resulted in a higher Compton-scattering probability and consequently in a larger number of firing segments (i.e. those measuring a net charge). 
When the $\sigma_Z$ values are normalized to the nominal height of the segments (see Figure~\ref{fig:FoR}), the resulting ratios are rather constant except for the front and back segments, where an increase is observed. 
For the former, the approximately 30\% excess can be explained by the segment geometry, since the height of the front segments is not uniform (see Figure~\ref{fig:Error-map}). 
For the back segments, instead, the increase mostly affects the uncertainties evaluated for the events where the number of firing segments is $\geq 3$, which is in agreement with the hypothesis on the $\gamma$-ray source position. 

Figure~\ref{fig:Error-Firing} shows also the dependence of position uncertainties on the number of firing segments. 
When only one segment per crystal was measuring a net charge, the positions obtained via the bootstrapping procedure are unequivocally equal to those determined from the original trace. 
This can be related to the fact that the signals constituting the database of reference were simulated assuming that only one segment fired. 
Thus, for these events, the FoM manifests a minimum so deep that the small residuals are negligible with respect to the difference between two neighboring grid points. 
This hypothesis was tested by introducing a random jitter of few samples (up to $\pm~50$ ns) in the generated traces, which again led to no variations in the position fluctuation distributions. 
Another argument in its favor is the observed increase of the uncertainty with the number of firing segments. 
Indeed, a comparison of one-hit basis signals with experimental traces, which result from an overlap of different signals, yields residuals that are much larger than the statistical fluctuations or than the electronic noise, and comparable to the signals themselves. 
Excluding the case of only one firing segment, the relation between the number of firing segments ($n$) and the position resolution ($W_p \equiv$ FWHM $\approx 2.355 \cdot \sigma_i$) was studied with the empirical formula

\begin{equation}
    W_p = f_0 +f_1 \sqrt{n} \,\,\,\,\,\,\,\, (n \geq 2) \, .
    \label{eq:firing}
\end{equation}
The $f_0$ and $f_1$ parameters were obtained for the \textit{Adaptive}-grid results of Figure~\ref{fig:Error-Firing} and their values are summarized in Table~\ref{tab:firing}.

\begin{table}[h]
    \centering
    \caption{\label{tab:firing}Position resolution as a function of the number of firing segments. For each segment geometry and for the whole detector, the values of the standard deviation shown in Figure~\ref{fig:Error-Firing} are fitted with an empirical relation given by Equation~\ref{eq:firing}. }
    \begin{tabular}{c c c c c c}
    \hline
    \multirow{2}{*}{Segment} &   \multicolumn{2}{c}{${XY}$}   &  &   \multicolumn{2}{c}{$Z$} \\
    \cmidrule{2-3}  
    \cmidrule{5-6}
            &       $f_0$ [mm]  &   $f_1$ [mm]      &   &       $f_0$ [mm]  &   $f_1$ [mm]\\
    \hline
    1        &   3.09 (57)   &   4.22 (24)   &  &   1.27 (26)   &   1.41 (9)    \\
    2        &   3.74 (49)   &   4.07 (21)   &  &   0.68 (24)   &   2.03 (9)    \\
    3        &   4.59 (80)   &   4.03 (35)   &  &   1.55 (26)   &   2.12 (9)    \\
    4        &   2.85 (47)   &   5.04 (19)   &  &   1.51 (49)   &   2.52 (21)   \\
    5        &   4.52 (89)   &   4.14 (38)   &  &   1.60 (61)   &   2.24 (26)   \\
    6        &   2.85 (49)   &   4.99 (21)   &  &   2.17 (85)   &   2.83 (35)   \\
    Total    &   3.60 (42)   &   4.43 (19)   &  &   0.45 (26)   &   2.57 (12)   \\
    \hline
    \end{tabular}
\end{table}

Figure~\ref{fig:Error-Firing} presents also a comparison of the results obtained with the \textit{Adaptive}~\cite{bazzacco2004PSA} and the \textit{Full} grid search modes: in the former, the PSA is first performed on a sparse grid of positions with a second step on the finer grid around the best FoM found on the sparse grid, while in the latter the search involves all points in the finer grid, giving the absolute minimum of the FoM. 
The uncertainties estimated with the $\textit{Full}$ grid search are systematically lower than those obtained in the \textit{Adaptive} mode, but the observed difference is below $10\%$. 
This small improvement may be meaningless considering that, due to a larger number of investigated points of the grid, both the PSA and bootstrapping procedures are very time-consuming in the \textit{Full} mode, which takes around $10$ times longer than the \textit{Adaptive} approach. 
Since such difference between the results of the two approaches has been observed for all investigated properties of the PSA-position uncertainty, in the following only the results of the \textit{Adaptive} grid-search mode are presented. 

\begin{figure}[h]
    \centering
    \includegraphics[width=0.48\textwidth]{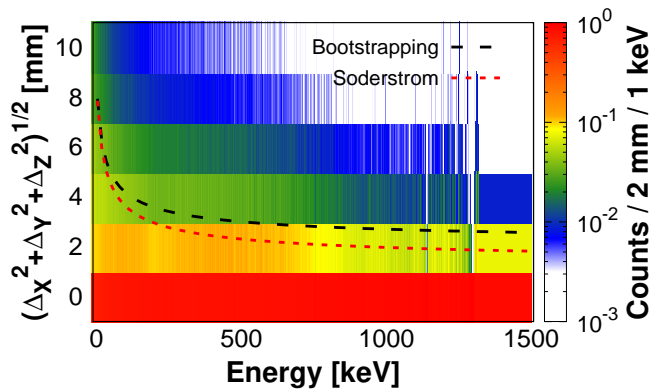}
    \caption{\label{fig:Error-Energy}(Color online) Position fluctuations as a function of the deposited energy. The curves represent the standard deviation as a function of energy, given by the bootstrapping technique (black dashed line) and the work of S\"{o}destr\"{o}m~\cite{soderstrom2011interaction} (red dotted line). Gamma-ray energies are limited to a range that is relevant for the present analysis, since the data were collected with a $^{60}$Co radioactive source. For each $\gamma$-ray energy the maximum intensity was normalized to 1 in order to simplify the comparison.} 
\end{figure}

\begin{figure*}
    \centering
    \includegraphics[width=0.42\textwidth]{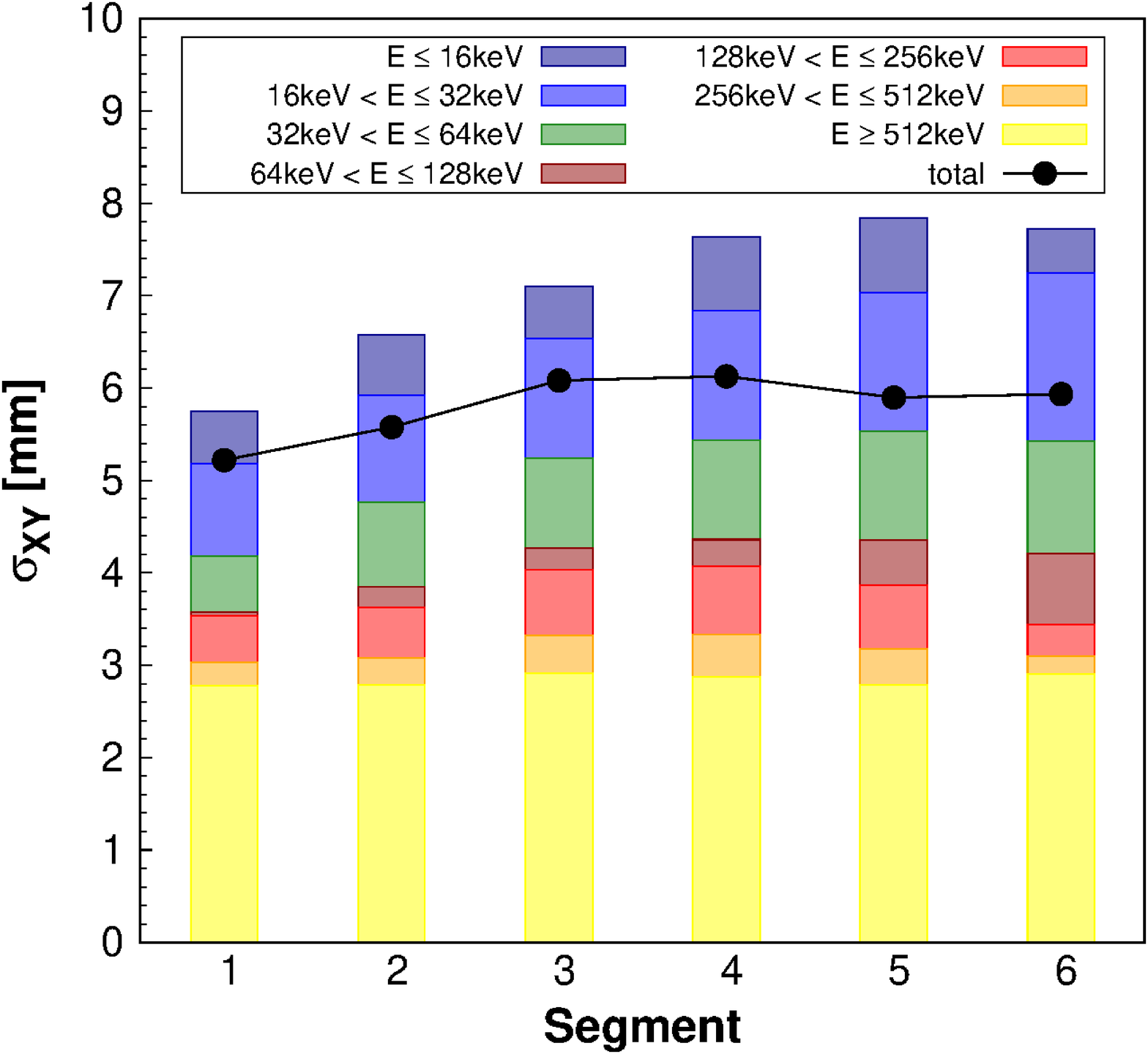}
    \hspace{5mm}
    \includegraphics[width=0.42\textwidth]{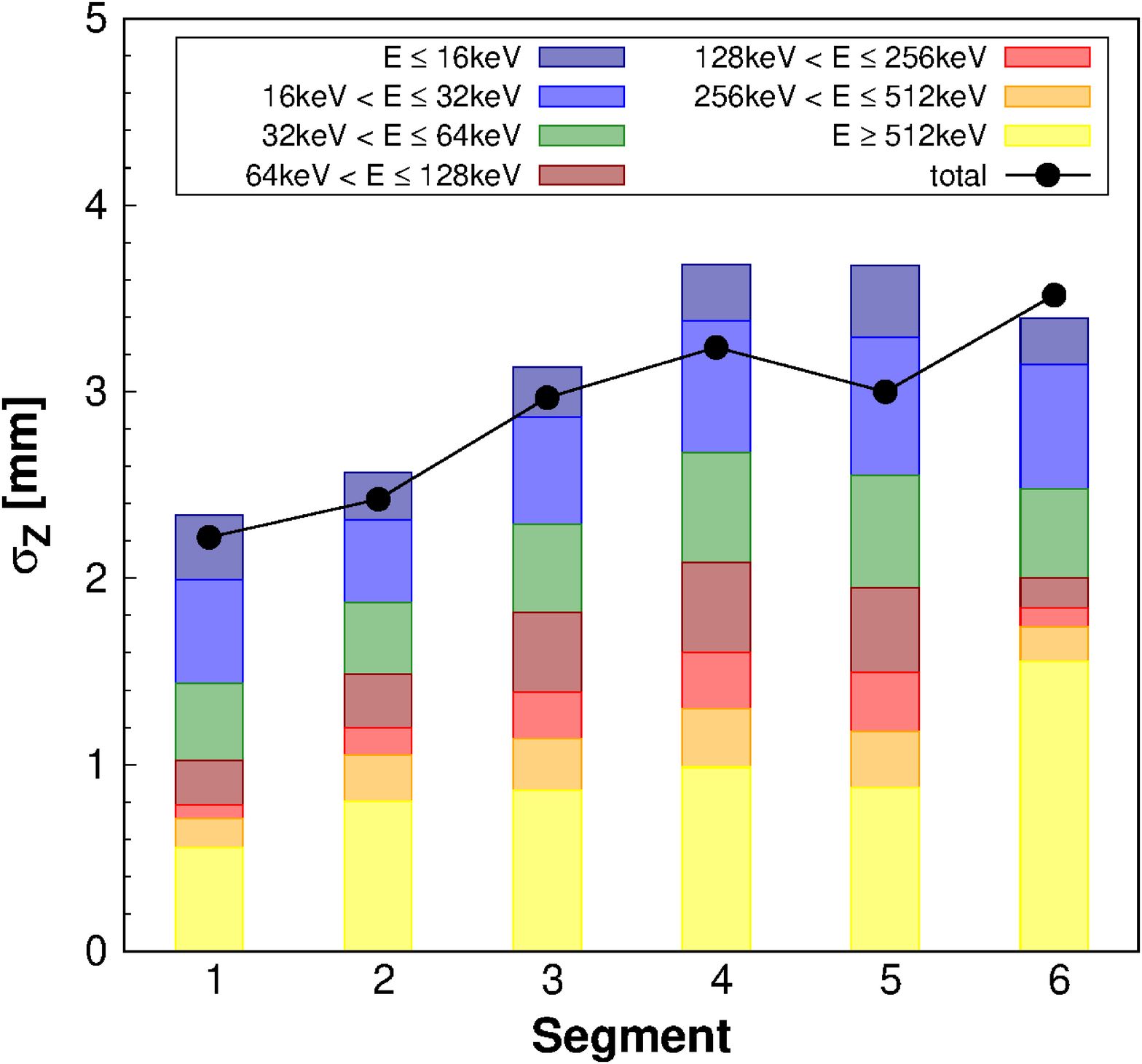}
    \vspace{-3mm}
    \caption{\label{fig:Error-EnergyXYZ}(Color online) Position uncertainties for the (left) radial and (right) $Z$ direction, for different energy ranges and segment geometries. The uncertainties are given by the standard deviation, assuming a symmetric distribution and using the \textit{Adaptive} grid search.}
\end{figure*}

As introduced before, several works have been already devoted to the study of the position resolution of the AGATA detector~\cite{RECCHIA200960,recchia2009position,soderstrom2011interaction,Steinbach2017,LJUNGVALL2020163297}. 
In particular, the study of S\"{o}der\hyp{}str\"{o}m and collaborators~\cite{soderstrom2011interaction}, performed by comparing Mon\hyp{}te-Carlo simulations with an in-beam experiment, discusses the dependence of the PSA uncertainty on energy. 
Developing a new method based on the energy resolution of Doppler-corrected $\gamma$-ray spectra, the authors proposed an empirical relation between the position resolution ($W_p$) and the deposited energy ($E_p$): 

\begin{equation}
    W_p (E_p) = w_0 +w_1 \sqrt{ \frac{100 \,\,\, keV}{E_p}} \, ,
    \label{eq:soderstrom}
\end{equation}
where the parameters $w_0 = 2.70 (17)$ mm and $w_1 = 6.2 (4)$ mm were extracted from the fit of various $\gamma$-ray transitions in the 200-4000 keV energy range. 

The dependence of the PSA uncertainty on the deposited energy has also been investigated via the bootstrapping technique. 
The evolution of the total position fluctuations $\Delta \equiv \sqrt{\Delta_X^2 +\Delta_Y^2 +\Delta_Z^2}$ was studied as a function of the deposited energy. 
As shown in Figure~\ref{fig:Error-Energy}, the standard deviation of the distribution rapidly decreases with increasing energy for E$_{\gamma} < 200$ keV and becomes asymptotically constant for higher energies. 
Since in the current study the position resolution displays the same behavior as the one given by the empirical relation of Ref.~\cite{soderstrom2011interaction}, the FWHM values obtained via the bootstrapping technique were fitted with Equation~\ref{eq:soderstrom}, yielding $w_0 = 4.63 (2)$ mm and $w_1 = 5.51 (3)$ mm. 
The obtained position resolution (FWHM$= 6.2$ mm at 1332.5 keV) is larger than the one deduced in Ref.~\cite{soderstrom2011interaction}. 
This difference can simply be due to specific properties of the investigated crystals, since in Ref.~\cite{LJUNGVALL2020163297} even larger systematic variations were observed between the individual detectors. 
Additionally, the method of Ref.~\cite{soderstrom2011interaction} requires the identification of specific $\gamma$-ray transitions emitted in-flight by the nucleus, and the results are systematically affected by the Doppler-correction and $\gamma$-ray tracking algorithms. 
In fact, for the position resolution deduced from the Doppler correction, the sensitivity to the interaction depth in the crystal is limited, as various depths correspond to the same $\gamma$-ray emission angle. 
In contrast, the bootstrapping approach can provide an almost continuous dependence of the position uncertainty on energy for each of the three directions, since it is only based on the deposited energy, independently of its origin. 

Figure~\ref{fig:Error-EnergyXYZ} presents the evolution of the position uncertainty for different energy ranges for the radial direction and along the $Z$ axis. 
The different energy ranges were selected as a compromise between the level of statistics needed to accurately investigate the deposited-energy dependence, and the required number of computational iterations. 
Also in this case the position resolution for the horizontal plane is rather constant for each energy range. 
Along the $Z$ axis, instead, while standard deviations remain constant for $E>256$ keV, for smaller energy depositions the position uncertainties increase with depth. 
This different behavior can be tracked down to the number of firing segments, since relatively small energy depositions in the bottom segments are mostly caused by the Compton scattering of energetic $\gamma$ rays.  

\begin{figure}[b]
    \centering
    \includegraphics[width=0.48\textwidth]{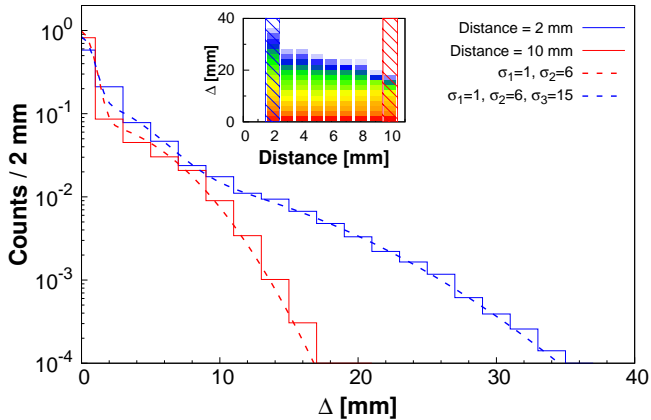}
    \caption{\label{fig:Error-Distance}(Color online) Distributions of total position fluctuations for the shortest (blue) and the longest (red) distance from the edge of a segment. The distribution integrals were normalized to 1 in order to simplify the comparison. The curves represent the functions used to fit the two distributions as described in the text: a sum of 3 Gaussian functions for the shortest distance (dashed blue line) and a sum of 2 Gaussian functions for the longest distance (dashed red line). In the inset, the fluctuations are presented as a function of the distance from the segment edge. }
\end{figure}

In order to understand the factors contributing to the position resolution and the role of segment geometry, the total position fluctuations $\Delta$ are analyzed as a function of the distance from the edge of the segments. 
Since the ADL bases are simulated assuming the nominal geometry of the HPGe crystals ~\cite{akkoyun2012agata,bruyneel2016ADL}, which may slightly differ from the real shape of the detectors, the precise position of the grid points inside the physical crystals is unknown. 
The distance from the border of the segments is, therefore, defined within the PSA-grid framework, making the minimum distance equal to the 2-mm grid step. 
Different behaviors are expected for interaction points near the borders and in the center of the segments, due to the fact that the FoM is calculated considering not only the net-charge signals but also the transient ones.
Figure~\ref{fig:Error-Distance} presents a comparison of $\Delta$ distributions obtained for the two extreme cases (i.e. segment edge and segment center). 
Both distributions can be well reproduced by a sum of Gaussian functions. 
Two Gaussians with $\sigma_1 = 1$ mm and $\sigma_2 = 6$ mm are necessary to describe the distribution for the points at the center of the segment, while for those at the border an additional long-range contribution with $\sigma_3 = 15$ mm is needed. 
While $\sigma_1$ is probably related to the size of the grid dimensions -- and, consequently, to the high statistics in the $\Delta = 0$ bin -- $\sigma_2= 6$ mm may be related to statistical fluctuations. 
Indeed, the sensitivity of the PSA is expected to be reduced close to the center of the segments, where the transient signals have the lowest amplitudes. 
The bootstrapping-induced fluctuations are probably similar to the signal fluctuations caused by electronic noise. 
For the positions close to the segments edge, instead, the importance of long-range fluctuations in the distribution is surprising. 
In fact, in the vicinity of the border, the shape of the pulses -- in particular of the transient-charge ones -- rapidly varies and the sensitivity of the PSA is very high. 
On one hand, such a behavior may have a geometrical origin and, as discussed in the following, be attributed to the very asymmetric shape of the distributions of $\Delta_X$, $\Delta_Y$ and $\Delta_Z$. 
On the other hand, it may be related to the ADL bases themselves.
In fact, it has recently been observed that when the distance between a PSA position and the virtual edge of the segment becomes lower than 0.5 mm, the shape of the simulated transient signal suddenly changes and starts to resemble that of the firing segment~\cite{Bart2018AGATAweek}. 
This clearly requires more investigation, but, if confirmed, could offer an explanation for the large residuals observed at segment borders. 
However, this result does not represent a failure of the bootstrapping approach. 
The technique, indeed, allows to infer the statistical properties of the PSA procedure as it is, pointing out possible systematic issues.

\begin{figure}
    \centering
    \includegraphics[width=0.48\textwidth]{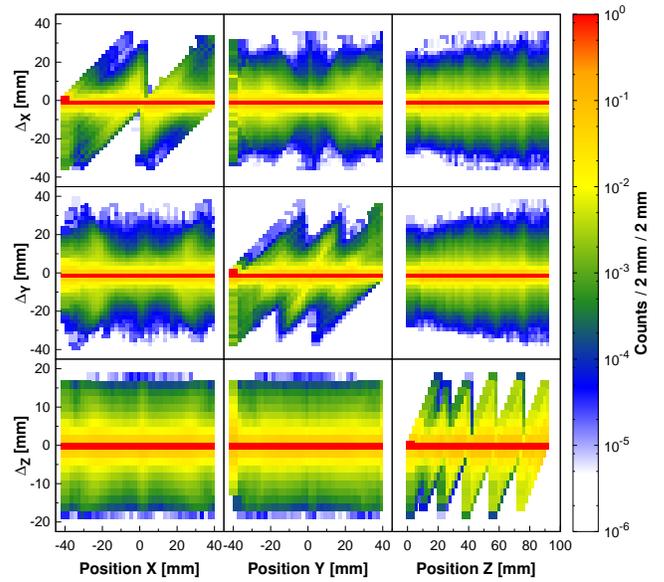}
    \caption{\label{fig:Error-Position}(Color online) Position fluctuations as a function of the PSA position, identified from the original experimental signals. The different patterns are caused by the geometry of the segments and their positions with respect to the frame of reference shown in Figure~\ref{fig:FoR}. For each position the maximum intensity was normalized to 1 in order to simplify the comparison.}
\end{figure}

\begin{figure*}
    \centering
    \includegraphics[width=0.96\textwidth]{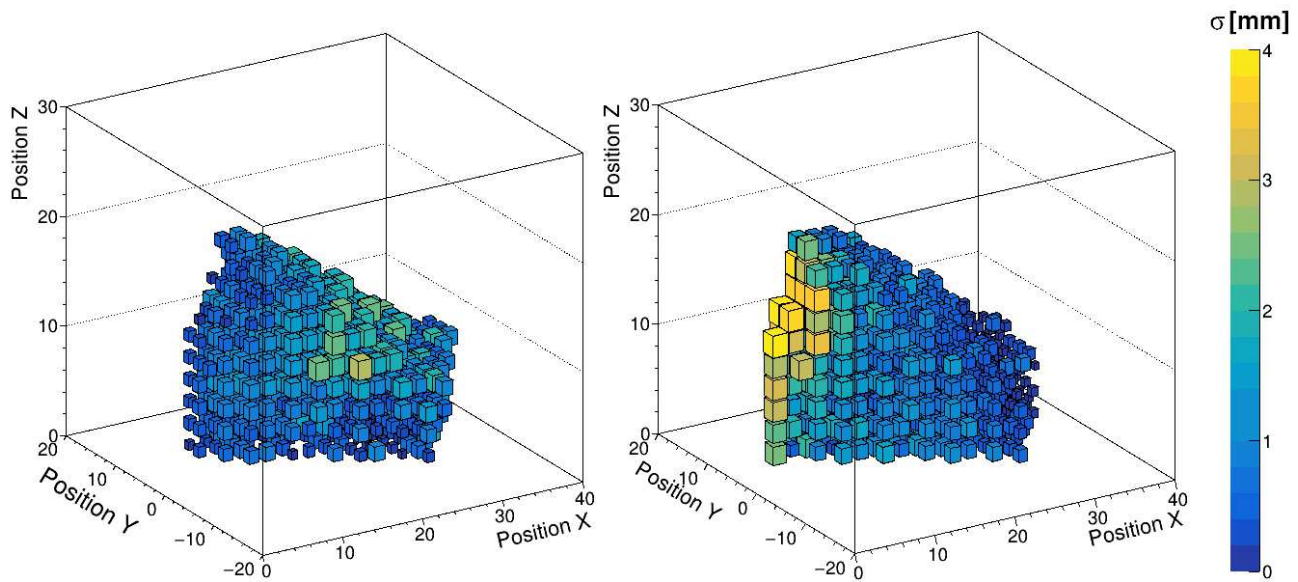}
    \caption{\label{fig:Error-map}(Color online) Position-uncertainty map for the $A1$ segment of the C013 AGATA detector. The lower (left) and upper (right) uncertainties along the $\vec{X}$ direction are estimated for the energy range 64 keV $< E_p \leq$ 128 keV. Colors and sizes of the boxes reflect the magnitude of the standard deviation.}
\end{figure*}

In order to study in more detail the effects of the interaction-point position on the spatial resolution, one can analyze how the PSA uncertainties evolve inside the segments of the AGATA detectors. 
Until now, the position resolution has been discussed assuming the dimensions of the segments as the only geometrical condition. 
However, since the firing segments are identified by the net charge, the uncertainties on the interaction-point positions cannot go outside of the segment boundary. 
This is, in fact, what can be observed in Figure~\ref{fig:Error-Position}, which presents how the $\Delta_X$, $\Delta_Y$ and $\Delta_Z$ position fluctuations depend on the PSA position of the original experimental signals. 
The different patterns reflect the geometry of the various segments, and in particular their position with respect to the internal frame of reference shown in Figure~\ref{fig:FoR}. 
For example, due to the symmetry of the detector and to the fact that the bootstrap position has to be within the firing-segment volume, the $\Delta_X$ fluctuations must be negative for positive values of $X$ and \textit{vice versa}. 
For the same reason, the evolution of the $\Delta_Z$ fluctuations as a function of the $Z$ coordinate reflects the vertical segmentation of the detector, with sign changes occurring at the segment borders. 

Therefore, as the fluctuation distributions are position-dependent, the position resolution cannot be treated as a general feature anymore, but it rather needs to be defined locally. 
Moreover, since the width of the fluctuation distribution rapidly varies with the distance from the segment border, asymmetric uncertainties have to be introduced. 
For each axis direction $k$, the lower and upper uncertainties were defined as the standard deviation of the negative and positive side of the respective distribution of $\Delta_k$, assuming $\vec{r}_i$ as the expected value (i.e. $\Delta_k=0$). 
This resulted in a multi-dimensional map of AGATA detectors, in which 6 energy-dependent values of uncertainty are provided for every grid position. 
An example of this map is shown in Figure~\ref{fig:Error-map} for the segment $A1$ of the C013 AGATA detector.

\section{Conclusions}
\label{sec:conclusions}

The possibility of performing pulse-shape analysis is an essential feature of new-generation HPGe arrays for high-resolution $\gamma$-ray spectroscopy studies.
A precise identification of the $\gamma$-ray interaction point is not only necessary for a more accurate Doppler correction, but it is also crucial when reconstructing the radiation path inside the detectors using tracking algorithms.
In this context, an evaluation of the position uncertainty provides a possibility to improve the tracking algorithms by assigning proper weights to the identified positions. 

In the present paper, the position resolution of AGATA detectors was studied via the bootstrapping technique. 
Despite being demanding in terms of computational resources and time, the simplicity of this statistical method allows to investigate the properties of the position uncertainty using practically any data sets (e.g. in-beam data, radioactive sources, background, etc.). 
Moreover, it permits to infer statistical features of a given hypothesis, in this case the whole PSA procedure, independently from external assumptions.

The dependence of the position resolution on various variables was analyzed. 
In particular, the relation between the position uncertainty and the deposited energy, observed previously in a study using the Doppler-correction method, has been confirmed. 
Since the PSA compares single-hit traces and experimental overlapping signals, the position uncertainty increases with the number of firing segments. 
Surprisingly, for the events when only one segment per crystal measures a net signal, the PSA result is identical for the original signals and for those generated using the bootstrapping technique. 
Due to the dimensions of the adopted grid, the FoM presents a very deep minimum, which is not affected by the fluctuations introduced via the bootstrapping method or by an eventual jitter of the signals. 
While investigating the relation between the position fluctuations and the distance from the edge of the segment, an unexpected behavior was observed for the positions at the segment border. 
This may be traced back to the pulses of the ADL bases, but it would require a dedicated study. 
The geometry of the segments and, in particular, the interaction-point position were shown to play a crucial role in the position resolution. 
As supposed in previous works, not only the position fluctuations do not present a Gaussian distribution, but their distributions are position-dependent and they exhibit an important asymmetry for PSA positions near the segment border. 
This work has been concluded by mapping AGATA detectors in terms of energy-dependent asymmetric standard deviations of interaction positions.

Further work is necessary to extend the study of the deposited-energy dependence to higher energies, in order to clearly observe the effects of the pair-production mechanism in the formation of the experimental signals. 
Moreover, in order to avoid the systematic bias related to the number of firing segments and to study the general properties of AGATA detectors, the present study should be repeated with a radiation source placed not only in front of the detector -- standard condition, similar to the in-beam measurements -- but also at its bottom. 

A natural continuation of this work would be the development of $\gamma$-ray tracking algorithms that use the information on position resolution to attribute weights to individual interaction points. 
Additionally, taking advantage of the information provided by the map of the position uncertainties, the future PSA basis for AGATA can be defined on an irregular-geometry grid. 
Such a solution, already partially adopted for the GRETINA/GRETA detectors~\cite{korichi2019}, allows reaching higher position sensitivity in certain detector regions, while limiting the ``useless'' iterations where the position resolution cannot be improved. 
On the other hand, the demonstrated dependencies of the position resolution may provide an impulse to study different segmentation schemes for the next generation of HPGe detectors.

\section*{Acknowledgement}
The authors would like to thank the AGATA collaboration and the GANIL technical staff. 
The excellent performance of the AGATA detectors is assured by the AGATA Detector Working Group. 
The AGATA project is supported in France by the CNRS and the CEA. 
This work has been supported by the OASIS project no. ANR-17-CE31-0026.
A.G. acknowledges the support of the Fondazione Cassa di Risparmio Padova e Rovigo under the project CONPHYT, starting grant in 2017. 
The authors are also grateful to CloudVeneto~\cite{cloud} for the use of computing and storage facilities.


\begin{thebibliography}{}
\bibitem{akkoyun2012agata}
	S.~Akkoyun \textit{et al.}, Nucl. Instr. and Meth. A \textbf{668}, (2012) 26.

\bibitem{paschalis2013GRETA}
    S.~Paschalis \textit{et al.}, Nucl. Instr. and Meth. A \textbf{709}, (2013) 44.

\bibitem{AGATA2020whitebook}
    W.~Korten \textit{et al.}, Eur. Phys. J. A \textbf{56}, (2020) 137.
    
\bibitem{AGATA@LNL2011}
    A.~Gadea \textit{et al.}, Nucl. Instr. and Meth. A \textbf{654}, (2011) 88.
    
\bibitem{AGATA@GSI2014}
	N.~Pietralla \textit{et al.}, Eur. Phys. J.: Web Conf. \textbf{66}, (2014) 02083.

\bibitem{Lalovic2016258}
    N.~Lalovi\'{c} \textit{et al.}, Nucl. Instr. and Meth. A \textbf{806}, (2016) 258.
    
\bibitem{AGATA@GANIL2017}
    E.~Cl{\'e}ment \textit{et al.}, Nucl. Instr. and Meth. A \textbf{855}, (2017) 1.
    
\bibitem{bruyneel2016ADL}
    B.~Bruyneel \textit{et al.}, Eur. Phys. J. A \textbf{52}, (2016) 700.

\bibitem{BRUYNEEL2006764}
    B.~Bruyneel \textit{et al.}, Nucl. Instr. and Meth. A \textbf{569}, (2006) 764.

\bibitem{BRUYNEEL2009196}
    B.~Bruyneel \textit{et al.}, Nucl. Instr. and Meth. A \textbf{599}, (2009) 196.

\bibitem{BIRKENBACH2011176}
    B.~Birkenbach \textit{et al.},Nucl. Instr. and Meth. A \textbf{640}, (2011) 176.

\bibitem{BRUYNEEL201192}
    B.~Bruyneel \textit{et al.}, Nucl. Instr. and Meth. A \textbf{641}, (2011) 92.

\bibitem{wiens2013}
    A.~Wiens \textit{et al.}, Eur. Phys. J. A \textbf{49}, (2013) 47.
    
\bibitem{CRESPI2008440}
    F.C.L.~Crespi \textit{et al.}, Nucl. Instr. and Meth. A \textbf{593}, (2008) 440.

\bibitem{Dimmock2009characterisation}
    M.R.~Dimmock \textit{et al.}, IEEE Transactions on Nuclear Science \textbf{56}, (2009) 1593.

\bibitem{HA2013123}
    T.M.H.~Ha \textit{et al.}, Nucl. Instr. and Meth. A \textbf{697}, (2013) 123.

\bibitem{Goel2013}
    N.~Goel \textit{et al.}, Nucl. Instr. Meth. A \textbf{700}, (2013) 10.

\bibitem{Ginsz2015}
    M.~Ginsz \textit{et al.}, \textit{Proceeding of the 4th International Conference on Advancements in Nuclear Instrumentation Measurement Methods and their Applications (ANIMMA), Lisbon, 2015}, (2015) 1.

\bibitem{HERNANDEZPRIETO201698}
    A.~Hernandez-Prieto \textit{et al.}, Nucl. Instr. and Meth. A \textbf{823}, (2016) 98.
    
\bibitem{HABERMANN201724}
    T.~Habermann \textit{et al.}, Nucl. Instr. and Meth. A \textbf{873}, (2017) 24.

\bibitem{deCanditiis2020}
    B.~de~Canditiis \textit{et al.}, Eur. Phys. J. A \textbf{59}, (2020) 276.
    
\bibitem{PhysRevC.62.024614}
    P.~D\'esesquelles \textit{et al.}, Phys. Rev. C \textbf{62}, (2000) 024614.

\bibitem{DESESQUELLES2011324}
    P.~D\'esesquelles, Nucl. Instr. and Meth. A \textbf{654}, (2011) 324.

\bibitem{DESESQUELLES2013198}
    P.~D\'esesquelles \textit{et al.}, Nucl. Instr. and Meth. A \textbf{729}, (2013) 198.

\bibitem{Li2018}
    H.~Li \textit{et al.}, Eur. Phys. J. A \textbf{54}, (2018) 198.
    
\bibitem{RecchiPHD}
    F.~Recchia, Ph.D. Thesis, Universit{\`a} degli Studi di Padova, 2008.
    
\bibitem{lewandowski2019PSA}
    L.~Lewandowski \textit{et al.}, Eur. Phys. J. A \textbf{55}, (2019) 81.

\bibitem{RECCHIA200960}
    F.~Recchia \textit{et al.}, Nucl. Instr. and Meth. A \textbf{604}, (2009) 60.

\bibitem{recchia2009position}
    F.~Recchia \textit{et al.}, Nucl. Instr. and Meth. A \textbf{604}, (2009) 555.

\bibitem{soderstrom2011interaction}
    P.-A.~S{\"o}derstr{\"o}m \textit{et al.}, Nucl. Instr. and Meth. A \textbf{638}, (2011) 96.

\bibitem{Steinbach2017}
    T.~Steinbach \textit{et al.}, Eur. Phys. J. A \textbf{53}, (2017) 23.
    
\bibitem{LJUNGVALL2020163297}
    J.~Ljungvall \textit{et al.}, Nucl. Instr. and Meth. A \textbf{955}, (2020) 163297.

\bibitem{lopez2004gamma}
	A.~Lopez-Martens \textit{et al.}, Nucl. Instr. and Meth. A \textbf{533}, (2004) 454.
    
\bibitem{efron1986bootstrap}
    B.~Efron \textit{et al.}, Statist. Sci. \textbf{1}, (1986) 54.

\bibitem{varian2005bootstrap}
    H.~Varian, Math. J. \textbf{9}, (2005) 768.

\bibitem{DESCOVICH2005199}
    M.~Descovich, \textit{et al.}, Nucl. Instr. and Meth. A \textbf{545}, (2005) 199.
    
\bibitem{bazzacco2004PSA}
	R.~Venturelli \textit{et al.}, INFN-LNL Annual Report \textbf{204}, (2004) 220.
	
\bibitem{Bart2018AGATAweek}
    B.~de~Canditiis \textit{et al.}, \textit{Contribution at the 19$^{th}$ AGATA week and 3$^{rd}$ Position Sensitive Germanium Detectors and Application Workshop, Strasbourg, 2018}.
    
\bibitem{korichi2019} 
    A.~Korichi \textit{et al.}, Eur. Phys. J. A \textbf{55} (2019) 121.

\bibitem{cloud}
	P.~Andreetto \textit{et al.}, Eur. Phys. J.: Web Conf. \textbf{214}, (2019) 07010.

\end{thebibliography}
\end{document}